\documentclass[11pt]{article}

\let\ssection=\section
\renewcommand{\section}{\setcounter{equation}{0}\ssection}

\def\parag{\hfil\break} 
\def\kikezd{\parag\underbar}
\def\ccr{\cr\noalign{\medskip}}
\def\p{{\partial}}
\def\vb{{\vec b}}
\def\vE{{\vec E}}

\def\vD{{\vec D}}
\def\vx{{\vec x}}

\def\vj{{\vec{\jmath}}}
\def\vnabla{{\vec\nabla}}
\def\vA{{\vec A}}
\def\vP{{\vec P}}
\def\cP{{\cal P}}
\def\vG{{\vec G}}
\newcommand\half{{\scriptstyle{\frac{1}{2}}}}

\begin{document}

\setlength{\baselineskip}{16pt}

\title{Galilean noncommutative  gauge theory:
   symmetries \& vortices}

\author{
P.~A.~Horv\'athy\footnote{e-mail: horvathy@univ-tours.fr}
\\
Laboratoire de Math\'ematiques et de Physique Th\'eorique\\
Universit\'e de Tours\\
Parc de Grandmont\\
F-37 200 TOURS (France)
\\
L. Martina\footnote{e-mail: Luigi.Martina@le.infn.it}\\
Dipartimento di Fisica dell'Universit\`a
\\
and\\
Sezione INFN di Lecce. Via Arnesano, CP. 193\\
I-73 100 LECCE (Italy).
\\ and\\
P.~C.~Stichel\footnote{e-mail: pstichel@gmx.de}\\
An der Krebskuhle 21\\
D-33 619 BIELEFELD (Germany)
}

\date{\today}

\maketitle

\begin{abstract}
     Noncommutative Chern-Simons gauge theory coupled to nonrelativistic
     scalars or spinors is shown to admit the
     ``exotic'' two-parameter-centrally extended Galilean symmetry,
     realized in a unique way consistent with the Seiberg-Witten map.
     Nontopological spinor vortices and topological external-field
     vortices are constructed by reducing the problem to previously
     solved self-dual equations.
\end{abstract}
\vskip5mm
\noindent\texttt{hep-th/0306228}


\section{Introduction}

In \cite{Hagen} Hagen suggested
to consider a nonrelativistic scalar field minimally
coupled to a gauge field with Chern-Simons dynamics.
When a suitable self-interaction potential is added, the system
admits exact self-dual vortex solutions \cite{JP}.

Noncommutative field theory has attracted much recent
  attention \cite{Szabo}. It was found, for example,
  that the free scalar theory in $2+1$ dimensions
is symmetric not only w. r. t. the usual one-parameter
centrally extended Galilei and Schr\"odinger groups, but also
with respect to their ``exotic'' two-parameter central
extensions \cite{HMS}. The hallmark of exotic symmetry is that the
components of the conserved boost generators do not commute,
\begin{equation}
     \big\{G_{i}, G_{j}\big\}=
     \epsilon_{ij}\theta\!\int\!\vert\psi\vert^2 d^2x
     \label{exoticrel}
\end{equation}
where $\theta$ is the noncommutative parameter \cite{exotic}.

Some potentials break the Galilean symmetry
\cite{Baketal}, while others do not \cite{HMS}.

Commutative gauge theory can accomodate also
  a fourth-order self-interaction potential so that it remains
invariant w. r. t. the conformal (Schr\"odinger) extension of the Galilei
group. This allows one to prove, e. g.,
that, for the critical coupling, all finite-energy solutions of
the second order field equations are selfdual \cite{JP}.
The scalar field  can also be replaced by a spinor so that
the modified theory  still supports self-dual vortices  \cite{DHP}.

The noncommutative version of the nonrelativistic
scalar field + Chern-Simons gauge
field model was considered by Lozano, Moreno and
Schaposnik \cite{LMSCS}, and by Bak, Kim, Soh, and Yee \cite{BakCS},
who also find exact, nontopological vortex-like solutions
which generalize those in the commutative theory.
\goodbreak

Our paper consists of two parts.
First we extend the symmetry investigations of \cite{HMS}
to noncommutative Chern-Simons gauge theory coupled to
scalars and spinors.
The second part is devoted to the study of various vortex
solutions.
Such theories are physically important
for the Fractional Quantum Hall Effect \cite{HaKa}.
These vortices correspond in particular to Laughlin's
quasiparticles and quasiholes.

Our paper is organized as follows.
With hindsight to the noncommutative generalization to come,
in  Section \ref{CGT} we review, following
the gauge independent approach of \cite{CM}, some
aspects of commutative Chern Simons gauge theory.

Sections \ref{NCGT}-\ref{SW}-\ref{scale} deal with the symmetry
properties of the noncommutative theory.
We argue that boosts should act from the right, and show that
this is the only possibility which is consistent with the
Seiberg-Witten map \cite{SW}.
Scale invariance is broken.

The NC vortices of \cite{LMSCS, BakCS} and of \cite{JaMaWa,BakNO,LMSNO}
are shortly discussed in Section \ref{NCvort}.

In Section \ref{fermions}, we extend our results to fermions.
Spinors were studied in \cite{DHP} in a Kaluza-Klein-type framework;
here we present a rather more direct approach that is readily
generalized to the noncommutative theory.

Topological vortices in a constant
(electro)magnetic background, relevant for the
FQHE, are discussed in our final Section \ref{extfieldvort}.

In both cases, the new vortices are constructed by reducing
the problem to the self-duality equation solved before by others,
namely \cite{LMSCS, BakCS}
in the nontopological case, and
\cite{JaMaWa,BakNO,LMSNO} in the topological case.
\goodbreak

\section{Galilean symmetry of commutative gauge theory}\label{CGT}

In the following we review briefly the main results of the
gauge-independent analysis presented in \cite{CM}.
One considers the Lagrangian
\begin{equation}
     L=L_{matter}+L_{field}=
     i\bar{\psi} D_{t}\psi-\half\big|{\vD\psi}\big|^2
     +\kappa\left(\half\epsilon_{ij}
     \p_{t}A_{i}A_{j}+A_{t}B\right)
     \label{Clag}
  \end{equation}
where $D_{\mu}=\p_{\mu}-ieA_{\mu}$ is the covariant derivative,
$B=\vnabla\times\vA$ and $E_{i}=\p_{i}A_{t}-\p_{t}A_{i}$.
The associated Euler-Lagrange equations consist
of a gauged Schr\"odinger equation,
(\ref{NLS}), of a field-current identity (FCI), (\ref{FCI}), and
of the constraint (\ref{Gauss}) which replaces the Gauss' law of Maxwell's
electromagnetism~:
\begin{eqnarray}
     iD_{t}\psi+\frac{1}{2}\vD^2\psi&=&0\label{NLS}
     \\[2pt]
    \kappa E_{i}-e\epsilon_{ik}j_{k}&=&0\label{FCI}
     \\[2pt]
     \kappa B+e\rho&=&0\label{Gauss}
\end{eqnarray}
The FCI is particularly relevant as it is precisely the Hall law.
  It follows that the density and the current,
\begin{equation}
\rho=\vert\psi\vert^2
\quad\hbox{and}\quad
j_{k}=\frac{1}{2i}\left((D_{k}\psi)\bar\psi-\psi(\overline{D_{k}\psi})\right)
\end{equation}
respectively, satisfy
the continuity equation $\p_{t}\rho+\vnabla\cdot\vj=0$.

Let us implement an infinitesimal
Galilean boost with parameter $\vb$ as
\begin{eqnarray}
     \delta^{0}_{}\psi&=&i\vb\cdot\vx\,\psi-t\vb\cdot\vnabla\psi
     \label{psiCimp}
     \\[2pt]
     \delta^{0}_{}A_{i}&=&-t\vb\cdot\vnabla A_{i}
     \label{ACimp}
     \\[2pt]
     \delta^{0}_{}A_{t}&=&-\vb\cdot\vA-t\vb\cdot\vnabla A_{t}.
     \label{A0Cimp}
\end{eqnarray}
Then, using the relations
\begin{equation}
     \begin{array}{lll}
     \delta^{0}_{}\rho=-t\vb\cdot\vnabla\rho,\qquad\hfill
     &\delta^{0}_{}{\vj}=-t\vb\cdot\vnabla\vj+\vb\rho,\hfill
     \\[3pt]
     \delta^{0}_{}B=-t\vb\cdot\vnabla B\hfill
     &\delta^{0}_{}E_{i}=-t\vb\cdot\vnabla E_{i} -\epsilon_{ij}b_{j}B\hfill
        \end{array}
        \label{relations}
\end{equation}
one proves readily that the system (\ref{NLS}-\ref{FCI}-\ref{Gauss}) 
is form-invariant
w. r. t. boosts. Positing the fundamental Poisson brackets
\begin{eqnarray}
     \big\{\psi(\vx,t),\bar{\psi}(\vx',t')\big\}&=&-i\delta(\vx-\vx')
     \label{psipsibar}
     \\[6pt]
     \big\{A_{i}(\vx,t),A_{j}(\vx',t')\big\}&=&\frac{\epsilon_{ij}}{\kappa}
     \delta(\vx-\vx')
     \label{Aij}
\end{eqnarray}
the field equations (\ref{NLS}-\ref{FCI}) can be recast
into a Hamiltonian form
$ 
\dot{Y}=\big\{Y,H\big\},
$ \ 
$Y=\psi,\bar\psi, A_{i}
$ 
  with the Hamiltonian
\begin{equation}
     H=\frac{1}{2}\int\!|\vD\psi|^2\, d^2\vx
     -\int\! A_{t}\big(e\rho+\kappa B\big)\, d^2\vx.
     \label{Hamiltonian}
\end{equation}

In restricted phase space defined by $A_{t}=\pi_{t}=0$ \cite{CM},
the momentum, the angular momentum, and the boosts have the form
\begin{eqnarray}
     P_{i}&=&\int\!\frac{1}{2i}
     \big(\bar{\psi}\p_{i}\psi-(\overline{\p_{i}\psi})\psi\big) d^2\vx
     -\frac{\kappa}{2}\int\!\epsilon_{jk}A_{k}\p_{i}A_{j} d^2\vx
     \label{momentum}
     \\[4pt]
     J&=&\int\!\left\{\epsilon_{ij}x_{i}\left(
     \frac{1}{2i}\big(\bar{\psi}\p_{j}\psi
     -(\overline{\p_{j}\psi})\psi\big)
     +\kappa\epsilon_{mn}A_{m}\p_{j}A_{n}\right)-\kappa A_{j}^2
     \right\}d^2\vx
     \label{angmom}\\[4pt]
     \vG^0&=&t\vP-\int\!\vx\,\rho\, d^2\vx.
     \label{cboost}
\end{eqnarray}
They are also constants of the motion. When constrained to the
surface defined by
the Gauss law (\ref{Gauss}), they assume more familiar forms,
\begin{equation}
     H=\frac{1}{2}\int\! |\vD\psi|^2,
\qquad
\vP=\int\!\vj,
\qquad
J=\int\!\vx\times\vj.
\end{equation}

Conversely,  an
infinitesimal coordinate change $\delta \vx$ is a symmetry
if it changes the Lagrangian by a surface term,
$\delta{\cal L}=\p_{\alpha}K^{\alpha}$. Then Noether's theorem
yields the constant of the motion
\begin{equation}
   \int\left(\frac{\delta{\cal L}}{\delta(\p_{t}\psi)}\delta\psi
  +\delta\bar{\psi}\frac{\delta{\cal L}}{\delta(\p_{t}\bar{\psi})}
     -K^{t}\right)d^2\vx
\label{Noether}
\end{equation}

Using the P.B. (\ref{psipsibar}) and
(\ref{Aij})  these quantities generate translations,
rotations, and boosts for the matter field and the
gauge field, respectively, according to
$ 
     -\p_{i}Y=\left\{Y,P_{i}\right\}
$ 
etc. 
The quantities above provide us with the usual
$1$-parameter centrally extended Galilei (``Bargmann'') algebra
with the particle number [mass], $M=\int\rho$, as central term.
In particular, the boost components commute,
\begin{equation}
     \big\{G_{1}^0,G_{2}^0\big\}=0.
     \label{comb}
\end{equation}
\goodbreak

\section{Noncommutative gauge theory}\label{NCGT}

Let us now turn to the noncommutative version of the above theory
\cite{LMSCS, BakCS}. The Lagrangian $L^l= L_{matter}^l+L_{field}^*$
is  {\it formally} the same as in the commutative case,
Eq. (\ref{Clag}); the noncommutative structure is
hidden in the definition of the covariant derivative and the field
strength,
\begin{eqnarray}
     D_{\mu}\psi=\p_{\mu}\psi-ieA_{\mu}\star\psi,\label{lcovder}
     \\
     F_{\mu\nu}=\p_{\mu}A_{\nu}-\p_{\nu}A_{\mu}
     -ie\big(A_{\mu}\star A_{\nu}-A_{\nu}\star A_{\mu}\big),
     \label{fieldstrength}
\end{eqnarray}
respectively, where the Moyal ``star'' product is associated
with the non-commutative parameter $\theta$,
\begin{equation}
\big(f\star g\big)(x_1, x_2)=\exp\left(i\frac{\theta}{2}\big(
\p_{x_1}\p_{y_2}-\p_{x_2}\p_{y_1}\big)\right)
f(x_1, x_2)g(y_1, y_2)\Big|_{\vx=\vec{y}}.
\label{thetaMoyal}
\end{equation}

According to (\ref{lcovder}) the matter field $\psi$
is in the fundamental representation of the gauge group
$U(1)_{*}$ i. e. $A_{\mu}$ acts from the left.
Hence $\overline{D_{\mu}\psi}=
\p_{\mu}\bar{\psi}+\bar\psi\star(ieA)$. Note also
that $L_{field}^*$ is equivalent to the Moyal-star Chern-Simons three-form
\begin{equation}
     \frac{\kappa}{2}\epsilon_{\mu\nu\sigma}
\big(A_{\mu}\star\p_{\nu}A_{\sigma}-\frac{2ie}{3}A_{\mu}\star A_{\nu}\star
A_{\sigma}\big).
\label{NCCSform}
\end{equation}

A remarkable feature of the NC
Chern-Simons theory is that gauge invariance requires
the  coefficient $\kappa$ to be
quantized even for the gauge group $U(1)_{*}$ \cite{kappaquant},
\begin{equation}
     \kappa=\frac{n}{2\pi},
     \qquad
     n=0,\pm1,\dots
     \label{quantcond}
\end{equation}

Apart from subtle differences, the field equations look
  as in the commutative case,
\begin{eqnarray}
     iD_{t}\psi+\frac{1}{2}{\vD}^2\psi&=&0\label{NCNLS}
     \\[2pt]
    \kappa E_{i}-{e}\epsilon_{ik}j^{l}_{\ k}&=&0\label{NCFCI}
     \\[2pt]
     \kappa B+e\rho^{l}&=&0\label{NCGauss}
\end{eqnarray}
where $B=\epsilon_{ij}F_{ij}$,  $E_{i}=F_{i0}$, and
$\rho^l$ and $\vj{\ }^l$ denote the {\it left density}
and {\it left current}, respectively,
\begin{eqnarray}
     \rho^{l}=\psi\star\bar\psi,
     \qquad
     {\vj}\strut{\,}^l=\frac{1}{2i}\left(\vD\psi\star\bar\psi
     -\psi\star(\overline{\vD\psi})\right).
     \label{ldenscur}
\end{eqnarray}
The condition (\ref{quantcond}) implies that
the Hall conductance is quantized in units of $(2\pi e)^{-1}$.

The continuity equation only holds for the
{\it right} quantities
\begin{equation}
     \rho^r=\bar{\psi}\star\psi,
     \qquad
{\vj}{\,}^r=\frac{1}{2i}\left(\bar\psi\star\vD\psi
     -(\overline{\vD\psi})\star\psi\right),
     \label{rdenscur}
\end{equation}
but not for the left-quantities (\ref{ldenscur}).
These latter satisfy in fact a covariant version, namely
$D_{t}\rho^l+\vD\cdot\vj{\,}^l=0$.
The integral property
implies $\int (\dot{\rho}^l+\vnabla\cdot\vj{\,}^l)=0$, though.

Owing to the ``handedness'',
the noncommutative system behaves somewhat unusually under
a Galilean boost.
Firstly, unlike in a pure scalar theory,
the field equations (\ref{NCNLS})-(\ref{NCFCI})-(\ref{NCGauss})
are {\it not} invariant w. r. t. the conventional boost
implementation
(\ref{psiCimp})-(\ref{ACimp})-(\ref{A0Cimp}). In fact,
  (\ref{relations}) and \# (2.1) of  \cite{HMS} imply that
\begin{equation}
\delta_{}^0B=-t\vb\cdot\vnabla B
\qquad\hbox{but}\qquad
\delta_{}^0\rho^{l}=-\frac{\theta}{2}\vb\times\vnabla\rho^l
-t\vb\cdot\vnabla\rho^l.
\label{convdenschange}
\end{equation}

The Gauss constraint is hence {\it not} form-invariant.

In \cite{HMS} we proposed another implementation which takes into account
the Moyal structure, i. e., to replace the above formulae by
the fundamental Moyal representation
\begin{equation}
     \delta^{l}_{}\psi=(i\vb\cdot\vx)\star\psi-t\vb\cdot\vnabla\psi
     =(i\vb\cdot\vx)\psi-(\theta/2)\vb\times\vnabla\psi
     -t\vb\cdot\vnabla\psi.
     \label{fimp}
\end{equation}
This still leaves the free theory invariant, and can also accomodate
  ``pure interactions''of the form $V(\rho^a)$ $a=l,r$.

But how to implement a boost on the gauge field ?
According to Eqn. (2.4) of \cite{JP02},
a coordinate transformation $f=f^{\alpha}$
which is at most linear in $\vx$ should be implemented as
\begin{eqnarray*}
\delta_fA_\mu=\half\big(f^\alpha\star\p_\alpha A_\mu
+\p_\alpha A_\mu\star f^\alpha\big)
+\p_\mu f^\alpha A_\alpha.
\end{eqnarray*}
But  $f_t=0, f_i=-tb_i$ for a boost, so this simply reduces to
the standard Lie derivative, $L_fA_\mu$, the same as
in the commutative case.
In conclusion, the standard implementation,
(\ref{ACimp})-(\ref{A0Cimp}),
is retained,
$\delta^lA_{\mu}=\delta^0A_{\mu}$. Hence
$\delta^lB=\delta_{}^0B=-t\vb\cdot\vnabla B$ as before.

However, as $\delta^l_{}\psi$ is in fact
$\delta^0_{}\psi-(\theta/2)\vb\times\vnabla\psi$
the $\theta$ terms add up making things ``even worse'',
\begin{equation}
\delta^{l}_{}\rho^{l}=
-\theta\vb\times\vnabla\rho^{l}-t\vb\cdot\vnabla\rho^{l}
\label{fundchangedens}
\end{equation}
cf. (\ref{convdenschange}). Hence, also the Moyal
implementation $\delta^l_{}$
breaks down for the gauged system~!
Galilean symmetry is restored, though, if we consider instead
the {\it antifundamental representation}
\begin{equation}
     \delta^{r}\psi=\psi\star(i\vb\cdot\vx)-t\vb\cdot\vnabla\psi
     =
     (i\vb\cdot\vx)\psi+\frac{\theta}{2}\vb\times\vnabla\psi
     -t\vb\cdot\vnabla\psi
     \label{afimp}
\end{equation}
which is (\ref{fimp}) with the sign of $\theta$ reversed.
Observing that
\begin{equation}
     \delta^r_{}\psi=\delta^0_{}\psi+\frac{\theta}{2}\vb\times\vnabla\psi
     \label{r0imp}
\end{equation}
we find that the $\theta$-terms cancel in $\delta^r\rho^l$,
leaving us with the homogeneous transformation law
\begin{equation}
     \delta^{r}\rho^{l}=-t\vb\cdot\vnabla\rho^{l}.
     \label{afundchangedens}
\end{equation}
Putting
$
\delta^{r}A_{\mu}=\delta^{0}_{}A_{\mu},
$
so that $\delta^rB=\delta^0B$, the
Gauss constraint (\ref{NCGauss}) is right-invariant,
just like the remaining equations.
For Eqn. (\ref{NCFCI}) this follows from
$$
\delta^rj^l_{k}=-t\vb\cdot\vnabla j^l_{k}+b_{k}\rho^l
\quad\hbox{and}\quad
\delta^rE_{i}=-t\vb\cdot\vnabla E_{i}-\epsilon_{ij}b_{j}B,
$$
while for (\ref{NCNLS}) this comes from
\begin{eqnarray*}
     \delta^r_{}\big(iD_{t}\psi+\half\vD^2\psi\big)=
     -t\vb\cdot\vnabla\big(iD_{t}\psi+\half\vD^2\psi\big)
     +\big(iD_{t}\psi+\half\vD^2\psi\big)\star (i\vb\cdot\vx).
\end{eqnarray*}

In conclusion, the antifundamental implementation (\ref{afimp})
allows us to restore the Galilean symmetry of the model.

The field equations can still be put into a Hamiltonian form,
using the same Poisson structure (\ref{psipsibar}-\ref{Aij}) as before.
When restricted to the surface in phase space defined by the Gauss' law
(\ref{NCGauss}),
the momentum, (\ref{momentum}), remains a constant of the motion.
For the boost generator we get, instead of
(\ref{cboost}),
\begin{equation}
     \vG^r=t\vP-\int\!\vx\rho^r\, d^2\vx
     \label{rncboost}
\end{equation}
whose conservation can also be checked directly, using the continuity
equation satisfied by $\rho^r$.
For the sake of comparision, we also  present the second term here as
\begin{equation}
     -\int\! x_{i}\vert\psi\vert^2\, d^2\vx
     -\frac{\theta}{2}\epsilon_{ij}\,
     \int\!\frac{1}{2i}\left(\bar{\psi}\p_{j}\psi-
     (\overline{\p_{j}\psi})\psi\right) d^2\vx\,
     \label{exorboost}
\end{equation}
which differs in the sign of $\theta$
from the analogous expression
for noncommutative scalar field theory,
\# (2.10) of \cite{HMS}.
This is due to our using the antifundamental,
rather than the fundamental representation.
Finally, for the commutator of the boost components we find
\begin{equation}
     \big\{G_{i}, G_{j}\big\}=\epsilon_{ij}k,
     \qquad
     k\equiv
     -\theta\!\int\!\vert\psi\vert^2 d^2x
     \label{exoticcharge}
\end{equation}
which  is (\ref{exoticrel}) with the sign of $\theta$ reversed.
  Apart from this, the two-parameter ``exotic'' central extension
\cite{exotic} is recovered. Our results extend those
obtained in \cite{HMS} to noncommutative CS gauge theory.

\section{Family of boost generators and the Seiberg-Witten (SW)
map}\label{SW}

One may wonder whether the boost generator given by (\ref{rncboost})
is unique.  In the free case it has been noted by Hagen \cite{Hagen}
that the conventional boost generator may be redefined by adding
$(\kappa/2)\epsilon_{ij}P_{j}$, which leads to a trivial second central
extension of the planar Galilei group. At first sight, the same seems to
hold also in noncommutative theory. One can indeed
define a whole family of generalizations
of the boost generator (\ref{rncboost}) depending on a real parameter
$\alpha$,
\begin{equation}
G_{i}^\alpha=G_{i}^r+\frac{\alpha}{2}\epsilon_{ij}P_{j}.
\label{genboost}
\end{equation}

Being a combination of two separately conserved quantities, $G_{i}^\alpha$
is plainly conserved, and
leads to the new transformation rules
\begin{eqnarray}
     \delta^\alpha\psi=\vb\cdot\big\{\psi,\vG^{\alpha}\big\}
     =
     i\vb\cdot\vx\,\psi-t\vb\cdot\vnabla\psi
     +\half(\theta-\alpha)\vb\times\vnabla\psi,
     \label{psialphaimp}
     \\[8pt]
     \delta^\alpha \vA=b_{i}\big\{\vA,G_{i}^{\alpha}\big\}
     =
     -t\vb\cdot\vnabla\vA-\frac{\alpha}{2}\vb\times\vnabla\vA.
     \label{Aalphaimp}
\end{eqnarray}

Then the Poisson brackets of the boost components would change from
(\ref{exoticcharge}) to
\begin{equation}
     \big\{G_{i}^{\alpha}, G_{j}^{\alpha}\big\}=
     \epsilon_{ij}(\alpha-\theta)\!\int\!\vert\psi\vert^2 d^2x
     \label{exoalphacharge}.
\end{equation}

For $\alpha=0$ we recover the right-implementation (\ref{afimp});
for $\alpha=\theta$ instead, we act on the matter field as in the
commutative case, (\ref{psiCimp}), but non-conventionally on the 
gauge potential;
we get a vanishing second  central charge.
The question arises, however, if the generalization
(\ref{genboost}) is allowed if we insist, in the spirit of Seiberg and Witten
\cite{SW} to recover the conventional implementation 
(\ref{psiCimp})-(\ref{ACimp})
in the commutative limit.
The matter and gauge fields in the noncomutative
($\theta\neq0$)
and commutative ($\theta=0$) theory must be indeed related
to each other by the Seiberg-Witten (SW) map \cite{SW}. In particular,
$\vA(\theta)$ must satisfy a differential equation \cite{SW}, Eq. (3.8)
namely
\begin{equation}
     \frac{\ \p}{\p\theta}A_{i}(\theta)=-\frac{1}{4}\epsilon_{kl}
     \Big(A_{k}\star (\p_{l}A_{i}+F_{li})+(\p_{l}A_{i}+F_{li})\star A_{k}\Big).
     \label{SWcond}
\end{equation}
Eqn.
(\ref{SWcond}) is manifestly form-invariant w. r. t. the boost transformations
(\ref{Aalphaimp}), provided $\alpha$ does not depend on $\theta$,
$
\frac{\ \p}{\p\theta}\alpha=0.
$
In the limit $\theta\to0$, this is consistent with the boost transformation
(\ref{ACimp}) only for $\alpha=0$.
In conclusion, the boost generator (\ref{rncboost}) is the only allowed one
if we require to recover the conventional implementation in the
commutative limit. The nontrivial second charge, (\ref{exoticcharge}), is
hence dynamically defined.

\section{Potentials and the breaking of the scale
invariance}\label{scale}

At this stage, we can add a potential. As explained in \cite{HMS},
the mixed expression $V(\rho^r\rho^l)$, favored, e. g., by Bak et al.
\cite{Baketal},
breaks the Galilean symmetry
whereas``pure'' expressions of the form
\begin{equation}
     V(\rho^r)\equiv V(\bar\psi\star\psi)
     \quad\hbox{or}\quad
     V(\rho^l)\equiv V(\psi\star\bar\psi)
     \label{purepot}
\end{equation}
are invariant w. r. t. both the conventional and the
``left-exotic'' [fundamental]
implementations, (\ref{psiCimp}) and $\delta^l_{}$, (\ref{fimp}),
respectively. Using (\ref{afundchangedens})
it is straightforward to prove that the same statement holds
for the antifundamental representation. In conclusion,
noncommutative CS theory augmented  with
  a pure potential $V(\rho^a)$ $a=r,l$
is consistent with (right-) Galilean symmetry.

As we said already, commutative Chern-Simons theory is consistent
with the ``Schr\"odinger'' (conformal) extension of the Galilei
group \cite{JP}, but any potential breaks
the scale invariance \cite{Baketal, HMS}. Let us now show that this
  also what happens in NC-CS gauge theory. Our proof relies on the
non form-invariant behaviour
of the Moyal product under a dilatation.
To see this, let consider a generic
star product $K_{1}\star K_{2}$ where the $K_{i}$ transform w. r. t.
an infinitesimal dilatation  as
\begin{equation}
     \delta_{\Delta}K_{i}=\Delta\cdot(k_{i}-\vx\cdot\vnabla-2t\p_{t})K_{i}
     \label{dilat}
\end{equation}
where $k_{i}$ is the scaling dimension and $\Delta\cdot$ means
multiplication with the parameter $\Delta>0$. Then it follows from
Eqn. (2.4) of \cite{HMS} that
\begin{equation}
     \delta_{\Delta}\cdot\big(K_{1}\star K_{2}\big)
     =\Delta\cdot\big(k_{1}+k_{2}-\vx\cdot\vnabla-2t\p_{t}\big)
     \big(K_{1}\star K_{2}\big)
     -i\theta\Delta\cdot\epsilon_{ij}\p_{i}K_{1}\star \p_{j}K_{2}
     \label{dilprod}
\end{equation}
where the  term behind $\theta$  breaks the form invariance.
This formula generalizes the one proved for the (right) density
in \cite{HMS}.

Let us now turn to the NC Gauss law (\ref{NCGauss}) which contains two
Moyal products, namely
the left-density, $\rho^l=\psi\star\bar{\psi}$, and the magnetic field,
$B$, which involves $i\epsilon_{ij}A_{i}\star A_{j}$. As the
individual factors transform as in (\ref{dilat}), the inhomogeneous
terms clearly break the scaling symmetry.

The same statement is readily seen to hold also for expansions.

\section{NC scalar vortices}\label{NCvort}

\subsection{Nontopological Chern-Simons vortices}\label{nontopCS}

In order to have a reference for the fermionic case (Section
\ref{fermions}), we review briefly the main results of
Lozano, Moreno, and Schaposnik \cite{LMSCS}, and of
Bak, Kim, Soh, and Yee \cite{BakCS}, respectively.
These authors consider  the previous NC-CS gauge theory to which
they add a quartic ``left-potential''
$V=(\lambda/4)\psi\star\bar{\psi}=(\lambda/4)\rho^l$.
This only changes (\ref{NCNLS}) into
\begin{equation}
     iD_{t}\psi+\frac{1}{2}{\vD}^2\psi+\frac{\lambda}{2}\rho^l\star\psi
     =0\label{vortNLS}
\end{equation}

The conserved quantities are routinely obtained.
For the energy we recover in particular their
  \begin{equation}
     H=\int\left(\frac{1}{2}\overline{\vD\psi}\vD\psi
     +\frac{\lambda}{4}(\rho^l)^2\right)d^2\vx.
     \label{NCenergy}
\end{equation}
Then, using the Bogomolny trick, for
$\lambda=\pm 2e^2/\kappa$ this becomes
\begin{equation}
     \int|D_{\pm}\psi|^2\,d^2\vx\geq0,
\end{equation}
where $D_{\pm}=D_{1}\pm iD_{2}$.
The Bogomolny bound, (namely zero) is therefore saturated
when the fields are self-dual or antiselfdual, respectively,
i. e. when
\begin{eqnarray}
	D_{\pm}\psi&=0\label{SD}
	\\[2pt]
	\kappa B+e\rho^l&=0\label{nontopGauss}
\end{eqnarray}

In the commutative case,  the upper equation (\ref{SD}) could be
solved for the vector potential; inserting the result into the lower one
would yield the Liouville equation with its known solutions \cite{JP}.
In the NC case, vortices are in turn constructed by
solving  these equations \cite{LMSCS, BakCS} using a rather involved
technique we do not reproduce here. We note for further
reference that
their SD ($D_{+}\psi=0$) solution is regular for $\kappa<0$,
and their ASD ($D_{-}\psi=0$) solution is regular for $\kappa>0$,
respectively. Their vortices are purely magnetic
as it can be seen from the second order field equations.
They are also
nontopological in that the density vanishes at infinity.

\subsection{NC Nielsen-Olesen vortices}\label{NCNO}

For the sake of their use in Section
\ref{extfieldvort}, we briefly review also
the noncommutative generalization of the
Nielsen-Olesen vortices examined in \cite{JaMaWa,BakNO,LMSNO}.
As this theory is relativistic, we will not review it in detail,
merely contend ourselves with mentionning that the static energy
functional can again be written using the Bogomolny trick as
\begin{equation}
     H=\int\left(\half\big(B\mp(\rho_{0}-\rho^l)\big)^2
     +(D_{\pm}\psi\overline{D_{\pm}\psi}\right)d^2\vx
     \pm\int\!\rho_{0}Bd^2\vx
\label{NObogdec}
\end{equation}
where $\rho_{0}>0$ is a constant.
The absolute minimum of the energy, namely
$\rho_{0}\big|(\hbox{magnetic flux})\big|$ is therefore attained when
the self-dual or the antiself-dual equations,
\begin{eqnarray}
     D_{+}\Phi=0\qquad&B=\rho_{0}-\rho^l\qquad
     &B>0\label{NOSD}
     \\[2pt]
     D_{-}\Phi=0\qquad&-B=\rho_{0}-\rho^l\qquad
     &B<0\label{NOASD}
\end{eqnarray}
hold, respectively.
In the commutative case the above procedure would yield a
``Liouville-type'' (but not explicitely soluble) equation;
in the NC case specific techniques were used \cite{BakNO, LMSNO}.

It is worth pointing out that here we follow, together with
Lozano et al. \cite{LMSCS, LMSNO}, the sign convention 
(\ref{thetaMoyal}). Some people
including Bak et al. \cite{BakNO} use the opposite sign for
$\theta$. Their results are, therefore, translated by interchanging
the words ``self-dual'' and ``antiself-dual'' \cite{LMSNO}. This
statement is not entirely obvious but can
be proved using the proporties of the Moyal product.
\goodbreak

\section{Fermions}\label{fermions}

\subsection{The gauged L\'evy-Leblond + Chern- Simons equations}

A Galilean covariant ``non-relativistic
Dirac equation'' was constructed by L\'evy-Leblond \cite{LLCMP}.
A Dirac spinor in $3+1$ dimensions has four-components, but
in the plane it only has two components; there are instead
two sets of ``Dirac'' matrices, appropriate to accomodate
spin $+\half$ and spin $-\half$.
(In \cite{DHP} the same two systems were obtained as the chiral
components of the $4$-component theory).
  We consider, for definiteness, the spin $\half$ theory.
Let hence $\Psi$ denote  a two-component spinor and
consider the fermionic matter Lagrange density
\begin{equation}
     L_{matter}=i\left\{\Psi^\dagger
     \big(\Sigma_{t}D_{t}-\vec{\Sigma}\cdot\vD
     -i\Sigma_{s}\big)\Psi\right\}
     \label{2LLlag}
\end{equation}
where the covariant derivatives have the same meaning as before, and
$\Sigma_{t}=\half(1+\sigma_{3}),$
     $\Sigma_{i}=\sigma_{i}\, (i=1,2),$
     $\Sigma_{s}=(1-\sigma_{3})$
denote the ``Dirac'' matrices. Observe that
$\Sigma_{t}$ and $\Sigma_{s}$ are singular
matrices.
Setting $\Psi=\left(\begin{array}{ll}\Phi\\
\chi\end{array}\right)$ yields the  first-order ``L\'evy-Leblond''
(LL) equations
\begin{equation}
     \begin{array}{ll}
	D_{+}\Phi+2i\chi=0
	\\[6pt]
	D_{t}\Phi-D_{-}\chi=0
	\label{LLeq}
     \end{array}
\end{equation}
where $D_{\pm}=D_{1}\pm iD_{2}$.
Choosing the same field Lagrangian as above,
we find that the Chern-Simons field equations
retain their form (\ref{FCI}) and (\ref{Gauss}) up to the definition
of the density and current,
\begin{equation}
     \rho=|\Phi|^2,
     \qquad
     \begin{array}{ll}
	j_{1}=-(\bar{\Phi}\chi+\bar{\chi}\Phi)
	\\[3pt]
	j_{2}=i(\bar{\Phi}\chi-\bar{\chi}\Phi)
	\end{array}
\label{spinordenscur}
\end{equation}
In particular, the density only involves
the upper component, since $\Sigma_{t}$ is a projector.

Implementing an infinitesimal boost as \cite{LLCMP, DHP}
\begin{equation}
     \begin{array}{ll}
     \delta^0_{}\Phi=(i\vb\cdot\vx)\Phi-t\vb\cdot\vnabla\Phi
     \\[6pt]
     \delta^0_{}\chi=-\half(b_{1}+ib_{2})\Phi
     +(i\vb\cdot\vx)\chi-t\vb\cdot\vnabla\chi\\
     \end{array}
     \label{spinCboost}
\end{equation}
while maintaining the previous implementation on the gauge fields,
we find that the LL equations (\ref{LLeq}) vary as\footnote{Due to
time translational symmetry, it is enough to vary at $t=0$.}
\begin{eqnarray*}
     i\vb\cdot\vx\Big(D_{+}\Phi+2i\chi\Big)
     \quad\hbox{and}\quad
     i\vb\cdot\vx\Big(D_{t}\Phi-D_{-}\chi\Big)-\half\big(b_{1}-ib_{2}\big)
     \Big(D_{+}\Phi+2i\chi\Big),
\end{eqnarray*}
respectively, which both vanish together with the LL equations (\ref{LLeq}).
This establishes the Galilean symmetry for the LL equations.
To extend this statement to the coupled system, we observe
that the spinor density and current in (\ref{spinordenscur})
change precisely as $\rho$ and $\vj$ in (\ref{relations}).
The invariance of the Chern-Simons equations is  hence retained.
\goodbreak

For a solution of the Chern-Simons field equations
the associated conserved quantities, derived using Noether's theorem,
can be expressed in terms of the upper component alone. They
are \cite{DHP} the mass [particle number], $M=\int|\Phi|^2$, and
\begin{equation}
     \begin{array}{ccc}
	\vP=&\displaystyle\int\vec\cP
	\,d^2\!\vx
	\equiv
	\displaystyle\!\int\left(\frac{1}{2i}
\left(\bar{\Phi}\vD\Phi-\overline{(\vD\Phi)}\Phi\right)
\right)d^2\!\vx\qquad\hfill
  &\hbox{linear momentum}\hfill
\\[8pt]
J=&J_{orbital}+J_{spin}
=\displaystyle\int
\vec{x}\times\vec{\cP}\,d^2\!\vx
\;+\;
\frac{1}{2}M\qquad\hfill
&\hbox{angular momentum}\hfill
\ccr
\vG=&t\vP
-\displaystyle\int\!\vx\,|\Phi|^2\,d^2\!\vx
\hfill
&\hbox{boost}\hfill
\\[8pt]
H=&
\displaystyle\int\left\{
{1\over2}|\vD\Phi|^2
+\frac{e^2}{2\kappa}
|\Phi|^4
\right\}\,d^2\!\vx\;
\hfill
\hfill&\hbox{energy}\hfill
\end{array}
\label{spinconserved}
\end{equation}
which shows clearly that the spin is indeed $\half$.
The components of the boost plainly commute,
(\ref{comb});  (\ref{spinconserved})
provides us with  the usual one-parameter centrally extended
Galilean relations \cite{LLCMP}.
\goodbreak

Then the usual trick applied to the energy
yields the Bogomolny bound
\begin{equation}
     H=\half\int\big|D_{+}\Phi\big|^2\geq0.
     \label{Bogdecomp}
\end{equation}

The absolute minimum of the energy
is attained therefore when the self-duality condition,
$D_{+}\Phi=0$ holds.
(Note that antiself-duality,
$D_{-}\Phi=0$, does {\it not} qualify here; it would be appropriate for
spin $-\half$).
Then exact, purely magnetic
solution can be constructed by solving again the Liouville equation
\cite{DHP}. Normalizable solutions arise provided $\kappa<0$.
An alternative, more detailed discussion will be presented
below in the NC context.

On the surface defined by the Gauss constraint (\ref{Gauss})
$H$ acts as a Hamiltonian.

\subsection{Noncommutative fermions}\label{NCferm}

The noncommutative generalization of these results is quite
strainghtforward. Both the matter and the field Lagrangian,
(\ref{2LLlag}) and (\ref{Clag}), retain their form, but the covariant
derivative and the field strength assume their NC meaning, cf. Section
\ref{NCGT}. The associated field equations are still (\ref{LLeq})
with the Moyal structure hidden in the covariant derivate, augmented
with the NC Chern-Simons equations (\ref{NCFCI}) and (\ref{NCGauss})
with left-density and current,
\begin{equation}
     \rho^l=\Phi\star\bar{\Phi},
     \qquad
     \begin{array}{ll}
	j\strut{\,}^l_{1}=-(\chi\star\bar{\Phi}+\Phi\star\bar{\chi})
	\\[6pt]
	j\strut{\,}^l_{2}=i(\chi\star\bar{\Phi}-\Phi\star\bar{\chi})
\end{array}
     \label{lNCdenscur}
\end{equation}

Galilean boosts have to be implemented by combining the right Moyal
action of the $\vx$-dependent imaginary factor, (\ref{afimp}),
with the spinor term in (\ref{spinCboost}). This yields simply
\begin{equation}
     \delta^r_{}\Psi=\delta^0_{}\Psi+\frac{\theta}{2}\vb\times\vnabla\Psi
     \label{spinNCboost}
\end{equation}
cf. (\ref{r0imp}). It follows  that (\ref{spinNCboost}) is again
a symmetry.
Firstly, the (NC) LL equation merely changes by
$
(\theta/2)\vb\times\vnabla\Big(\ \hbox{LL eqn}\ \Big).
$
As the  density and current change once again as before, the
boost invariance of the NC spinor system is established.

It follows from (\ref{spinNCboost}) that
the associated conserved quantities are obtained by combining those
in the NC scalar case with the commutative spinorial expressions in
(\ref{spinconserved}). For a boost, e. g., Eq. (\ref{rncboost}) is
still valid, when $\psi$ is replaced by the upper component $\Phi$.
The boost components satisfy the exotic relation (\ref{exoticcharge}).
Similarly, the energy is
\begin{equation}
     H=\int\left(\frac{1}{2}\overline{\vD\Phi}\vD\Phi
     +\frac{e^2}{2\kappa}(\rho^l)^2\right)d^2\vx,
     \label{NCspinenergy}
\end{equation}
which is precisely the same as in the NC scalar theory studied by
Lozano et al. \cite{LMSCS}, and could be used therefore
to construct vortex solutions.

We prefer, however, to follow another procedure
which is peculiar to the first-order setting.
Let us observe indeed that the static and purely magnetic Ansatz
\begin{equation}
     D_{+}\Phi=0,\qquad
     \chi=0,\qquad
     \p_{t}\Phi=0,\qquad
     A_{t}=0
     \label{spinsol}
\end{equation}
plainly solves the static version of the LL equations (\ref{LLeq}).
Then both the electric field, $\vec{E}$, and the current, $\vj{\ }^{l}$
vanish cf. (\ref{spinordenscur}), so that the
FCI (\ref{NCFCI}) holds identically, and we are left with the
Gauss law (\ref{Gauss}). As $\rho^l=\Phi\star\bar{\Phi}$,
  we arrive at the problem solved before by Lozano et al.,
  and by Bak et al. \cite{LMSCS, BakCS}.
The only difference is that our spinor vortices
are purely magnetic, while those in \cite{LMSCS} carry also an electric field
-- as do their commutative limit \cite{JP}.
This ``coincidence'' is explained by
eliminating the ``lower'' component $\chi$ from the LL equation
by using the identity
$$
D_{-}D_{+}=\vD^2+i[D_{1},D_{2}]=\vD^2+eB\star.
$$
Then the ``upper'' component
satisfies the gauged Schr\"odinger equation with a Pauli term
\footnote{For spin $-\half$ one merely changes the sign of $\Sigma_{2}$
which results in changing the sign of the magnetic field in (\ref{SP}).},
\begin{equation}
     \left[iD_{t}+\half\vD^2+\half eB\star\right]\Phi=0
     \qquad\hbox{i. e.}\qquad
     \left[iD_{t}+\half\vD^2+\frac{e^2}{2\kappa}\rho^l\star\right]\Phi=0
     \label{SP}
\end{equation}
where the Gauss law (\ref{NCGauss}) has been used.
This is precisely the field equation of Lozano et al. \cite{LMSCS},
(\ref{vortNLS}),
with the specific SD value of the coupling constant. It can be
solved using the solutions of the SD equations (\ref{SD})-(\ref{nontopGauss}).
Note  that the quartic ``left-potential''  with the ``SD'' coupling
coefficient here
was {\it not} put in by hand but came rather from the (``left'') Gauss
law (\ref{NCGauss}).

The spinor theory is, in this sense, automatically self-dual.
\goodbreak

\section{Vortices in the Landau-Ginzburg theory of the QHE
}\label{extfieldvort}

A ``Landau-Ginzburg'' theory for the QHE has been
proposed by  Zhang {\it et al.\/} \cite{Zhang}.
They consider a scalar field $\psi$ coupled to a gauge field $a_{\mu}$
and an external electromagnetic potential $\tilde{A}_{\mu}^{ext}$,
described by the Lagrangian
\begin{equation}
\frac{\kappa}{2}\epsilon^{\mu\nu\sigma}a_{\mu}\p_{\nu}a_{\sigma}
+\bar{\psi}\big[i\p_{0}-e(a_{t}+\tilde{A}_{t}^{ext})\big]\psi
-\bar{\psi}\big[-i\vnabla-e(\vec{a}+\vec{\tilde{A}}^{ext})\big]^{2}\psi-U(\psi)
\label{ZHKlag}
\end{equation}
where $U(\psi)$ is a (quartic) self-interaction potential.
The gauge field $a_{\mu}$ obeys hence the Chern-Simons dynamics,
but the matter field $\psi$ moves under the joint influence of
$a_{\mu}$ and the external field.
Let us assume that the background field is constant.
Putting
$A_{t}=-a_{t}-A_{t}^{ext}$ and $\vec{A}=\vec{a}+\vec{A}^{ext}$,
performing some partial integrations and dropping surface terms,
  this is further  written as
\begin{equation}
\frac{\kappa}{2}\epsilon^{\mu\nu\lambda}A_{\mu}\p_{\nu}A_{\lambda}
     -e\left(B^{ext}A_{t}
-\vec{E}^{ext}\times\vec{A}\right)
+i\bar{\psi}D_{t}\psi-\half\big|D\psi\big|^2-U(\psi)
\label{Manton}
\end{equation}
where $D_{\mu}=\p_{\mu}-ieA_{\mu}\psi$ and the external field has
been redefined as
$F^{ext}_{\ \ \mu\nu}=(\kappa/e)\tilde{F}^{ext}_{\ \ \mu\nu}$.
Augmented with the natural magnetic term $B^2/2$, the system (\ref{Manton})
was introduced by Manton \cite{Manton}. Note also that the second,
``external'' term here is also equivalent to
$
-e\epsilon^{\mu\nu\lambda}A^{ext}_{\mu}\p_{\nu}A_{\lambda}
$
considered in \cite{HaKa}.

In \cite{Manton} Manton found in particular
exact purely magnetic vortices as
imbedded Bogomolny-Nielsen-Olesen solutions.
His results were generalized to vortices with an
electric field in \cite{HHY}.

Recently, it was argued that to describe
the Fractional Quantum Hall Effect the commutative Chern-Simons term
should be replaced by the noncommutative expression (\ref{NCCSform})
  \cite{HaKa}. Here we propose therefore to consider
the noncommutative model given by
\begin{equation}
     \begin{array}{cc}
     L^{ext}=&-\half B^2+i\bar{\psi}D_{t}\psi-\half|\vD\psi|^2
     -\frac{\lambda}{4}(\rho_{0}-\psi\star\bar{\psi})^2\hfill
     \\[12pt]
     &+\frac{\kappa}{2}\epsilon^{\mu\nu\lambda}
     \Big(A_{\mu}\star\p_{\nu}A_{\lambda}
     -\frac{2ie}{3}A_{\mu}\star A_{\nu}\star A_{\lambda}\Big)
     -e\left(B^{ext}A_{t}
     -\vec{E}^{ext}\times\vec{A}\right)\hfill\\
     \end{array}
     \label{NCMantonlag}
\end{equation}
where $\rho_{0}$ is a constant.
The covariant
derivative and the field strength are again as in
(\ref{lcovder})-(\ref{fieldstrength}).
The absence of an electric Maxwell term is dictated
by Galilean, rather than Lorentz covariance \cite{HHY}.
If the external fields are constant, the system
is  translational invariant.
The ``naked'' external-field term $-eB^{ext}A_{t}$ modifies
the quantization condition (\ref{quantcond}) as \cite{kappaquant}
\begin{equation}
     \kappa+eB^{ext}\theta=\frac{n}{2\pi},
     \qquad
     n=0,\pm1,\dots
     \label{backquantcond}
\end{equation}

$A_{t}$ is a Lagrange multiplier and variation w. r. t. it
yields the modified Gauss constraint
\begin{equation}
      \kappa B=e(B^{ext}-\rho^{l}).
      \label{modGauss}
\end{equation}

Before considering the remaining field equations
let us inquire about finite-energy configurations.
If the external electric field is also constant that we assume henceforth,
it can be eliminated by a Galilean boost.
In the frame where $\vE^{ext}=0$  the energy can be expressed as
\begin{equation}
     \int\left(\half|\vD\psi|^2
     +\half B^2
     +\frac{\lambda}{4}\big(\rho_{0}-\rho^l\big)^2
     \right)d^2\vx.
     \label{extfiener}
\end{equation}
This expression can be justified, e.g., by considering
the Hamiltonian associated with (\ref{NCMantonlag}) and
using the Gauss constraint.
Let us call attention to that both the electric
term $\half{\vE}^2$ and the time-derivative of the matter field
are absent here. This follows from the nonrelativistic form
(\ref{NCMantonlag}) of the action. This is in contrast with the
relativistic setting of Nielsen-Olesen, where these terms
are eliminated by a static, purely-magnetic Ansatz.

In order to make the energy integral converge, we require
\begin{equation}
     \vD\psi\to0,
     \qquad
     B\to0
     \qquad
     \rho^l\to\rho_{0}
     \label{finiteenergy}
\end{equation}
sufficently rapidly as $r=|\vx|\to\infty$. Comparision with
  (\ref{modGauss}) shows that necessarily
$B^{ext}=\rho_{0}>0$.

Let us now turn to the field equations,
\begin{eqnarray}
     iD_{t}\psi+\frac{1}{2}{\vD}^2\psi+
     \frac{\lambda}{2}(B^{ext}-\rho^l)\star\psi=0\label{NCExNLS}
     \\[4pt]
    \kappa\epsilon_{ik}E_{k}
    +e(j^{l}_{\ i}+\epsilon_{ik}E^{ext}_{k})
    -\epsilon_{ik}D_{k}B=0\label{Ampere}
     \\[4pt]
     \kappa B=e(B^{ext}-\rho^{l})\label{NCExGauss}
\end{eqnarray}

The first equation here is a non-linear Schr\"odinger equation
which involves the left-density $\rho^l$; the second one combines
  Amp\`ere's law with the FCI appropriate for the Hall effect.
  (The current $j^{l}_{\ i}$ here is (\ref{ldenscur})).
The last is a modified Gauss law.

Now we show that for a suitable $\lambda$
this system admits self-dual vortex solutions.
Let us namely combine the self-duality Ansatz (\ref{SD}) with the
modified Gauss law (\ref{NCExGauss})
\begin{equation}
     \begin{array}{ll}
     D_{\pm}\psi=0\hfill
     \\[4pt]
     \kappa B=e(B^{ext}-\rho^{l}).\hfill
     \end{array}
     \label{topSD}
\end{equation}
In the frame $\vE^{ext}=0$,
the static version of the upper two field equations require,
using $\vD^2\psi=\mp eB\star\psi$,
\begin{eqnarray*}
    \left[eA_{t}\mp\frac{1}{2}eB+
     \frac{\lambda}{2}(B^{ext}-\rho^l)\right]\star\psi=0
     \\[3pt]
    \kappa\epsilon_{ik}E_{k}+ej^{l}_{\ i}-\epsilon_{ik}D_{k}B=0.
\end{eqnarray*}
Then the first equation is satisfied
with $A_{t}=\mu\big(B^{ext}-\rho^l\big),$
where $\mu=\half\left(\pm\frac{e}{\kappa}-\frac{\lambda}{e}\right).$
Using self-duality (\ref{topSD}), we find also
$$
\vj^{\ l}=\mp\half\vD\times\rho^l,
\qquad
\vE=-\mu\vD\rho^l,
\qquad\hbox{and}\qquad
\vD\times B=-\frac{e}{\kappa}\vD\times\rho^l.
$$
[Here $\vD$ acts on $\rho^l$ in the adjoint representation].
Then Amp\`ere's law  fixes the coefficient of the
self-interaction potential and hence the electric potential as
\begin{equation}
     \lambda=\frac{2e^2}{\kappa^2}(\pm\kappa-1)
     \qquad\Longrightarrow\qquad
     A_{t}=\frac{e}{2\kappa^2}(2\mp\kappa)\,(B^{ext}-\rho^l).
     \label{goodlambda}
\end{equation}

The self-interaction
potential is physically admissible (attractive) when $\lambda\geq0$.
For the upper and lower signs this requires $\kappa\geq1$ and
$\kappa\leq-1$, respectively. Interestingly, for
$
\kappa=\pm1,
$
the SD equations work with $\lambda=0$,
i. e., without a self-interaction potential. The asymptotic
behaviour $\rho^l\to B^{ext}$ is still mandatory, owing to
$B\to0$ and the Gauss law.

The self-duality equations (\ref{topSD})
could also have been derived using the Bogomolny trick.
Using the Gauss constraint and the identity
$$
|\vD\psi|^2\sim|D_{\pm}\psi|^2\pm eB\star\rho^l
$$
where ``$\sim$'' means up to surface terms,
the energy can be further written as
\begin{equation}
     \int\left(\half|D_{\pm}\psi|^2
     +\frac{1}{4}
     \Big(\lambda+\frac{2e^2}{\kappa^2}(1\mp\kappa)\Big)
     \big(B^{ext}-\rho^l\big)^2
     \right)d^2\vx
     \pm \half B^{ext}\int eB\,d^2\vx.
     \label{energydecomp}
\end{equation}

The last integral here is the magnetic flux,
$\int Bd^2\vx$.
For the specific choice (\ref{goodlambda}) of $\lambda$
the middle term vanishes and, chosing the upper/lower
sign depending on the sign of $eB$ yields
the usual Bogomolny inequality for the energy
\begin{equation}
     H\geq \half B^{ext}\Big|e\times (\hbox{magnetic flux})\Big|.
     \label{bogbound}
\end{equation}
Equality is achieved here for the SD/ASD equations (\ref{topSD}).

In the commutative context, the flux is quantized \cite{JaTa},
\begin{equation}
     \hbox{magnetic flux}=\int Bd^2\vx=\frac{2\pi}{e}\times n.
     \label{fluxquant}
\end{equation}
In the noncommutative theory the situation is less clear.
For topological vortices of the Nielsen-Olesen type,
circumstancial evidence \cite{JaMaWa, LMSNO} indicates that
(\ref{fluxquant}) likely remains true, even if no general
proof is available as yet. For comparision, for the
  nontopological vortices that appear in nonrelativistic  Chern-Simons
theory without an external field the flux
  does not appear to be quantized \cite{LMSCS}.

At this stage, self-dual solutions can be constructed
using those results in \cite{JaMaWa,BakNO,LMSNO}
\footnote{The fact that the latter investigations concern the
noncommutative
{\it relativistic} Abelian Higgs model are without importance
here:  we are interested by solving the {\it same} equations
with the {\it same} boundary conditions, and we can ignore
their origin.}.
Comparision with the Nielsen-Olesen SD equations (\ref{NOSD}-\ref{NOASD})
reveals, however, a subtle difference:
we don't have the freedom to chose the sign
in the second equation~: our (\ref{topSD}) is in fact
one of the field equations. There is instead the freedom in chosing
the sign of $\kappa/e$.
Redefining $B$ as $\tilde{B}=\big|\frac{\kappa}{e}\big|B$ brings
indeed (\ref{topSD})
to the SD form in (\ref{NOSD}) for $\frac{\kappa}{e}\geq0$,
  and to the ASD form (\ref{NOASD}) for $\frac{\kappa}{e}\leq0$.

Then the SD solutions constructed in \cite{JaMaWa,BakNO,LMSNO}
provide us with nonrelativistic, external-field vortices.
Our vortices carry a (statistical) electric field
except for $\kappa=\pm2$  while those,
relativistic, considered in
\cite{JaMaWa,BakNO,LMSNO} are purely magnetic.
Our  NC external-field vortices also
differ from those Maxwell-Chern-Simons
objects in \cite{Khare} as these latter are fully relativistic.

As a final example, let us  combine the
models in Sections \ref{fermions} and \ref{extfieldvort}
i. e., consider a spin $\half$ field
in a constant external field. In the frame where
$\vE^{ext}=0$, we study hence the static system
\begin{eqnarray}
     eA_{t}\star\Phi+\half\vD^2\Phi+\frac{e}{2}B\star\Phi=0\label{SchP}
     \\[4pt]
     \kappa\epsilon_{ik}E_{k}+ej_{i}^{\ l}-\epsilon_{ik}D_{k}B=0
     \label{AmpereFCI}
     \\[4pt]
     \kappa B=e\big(B^{ext}-\rho^l\big)
     \label{NCgaussbis}
\end{eqnarray}
cf. (\ref{SP}-\ref{Ampere}-\ref{NCExGauss}), where $\Phi$ denotes
the upper component of the Pauli spinor
$\left(\begin{array}{ll}
\Phi
\\
\chi
\end{array}\right)
$.
Eliminating the $\chi$
component, the current reads
\begin{equation}
     j_{1}^{\ l}=\frac{1}{2i}
     \big(D_{+}\Phi\star\bar{\Phi}-\Phi\star\overline{D_{+}\Phi}\big),
     \qquad
     j_{2}^{\ l}=
     -\frac{1}{2}
     \big(D_{+}\Phi\star\bar{\Phi}+\Phi\star\overline{D_{+}\Phi}\big).
     \label{spinorcurrents}
\end{equation}
The electric field is $\vE=\vD A_{t}$.

Let us now search for solutions.
As a first attempt, try the self-duality, $D_{+}\Phi=0$.
Then $\vD^2\Phi=-eB\star\Phi$ so that (\ref{SchP}) requires
$A_{t}=0$. But then $\vE=0$ and as plainly $\vj=0$, Amp\`ere's
law, (\ref{AmpereFCI}), only allows for a trival magnetic field,
$\vD B=0$. No SD solution is hence obtained.

Somewhat surprisingly, {\it antiselfdual} solutions may exist,
however \cite{HHY}. For $D_{-}\Phi=0$ we have instead
$\vD^2\Phi=+eB\star\Phi$
so that the static Schr\"odinger-Pauli equation (\ref{SchP})
[as well as its ``square root, the gauged L\'evy-leblond equation
(\ref{LLeq})]
can be satisfied with a nontrivial scalar potential, $A_{t}=-B$.
As now
$
D_{+}\Phi=2D_{1}\Phi=2iD_{2}\Phi,
$
the currents do not vanish but are rather
expressed as $j_{i}=\epsilon_{ik}D_{k}\rho^l$.  Amp\`ere's law
(\ref{AmpereFCI})  requires therefore
$
\big(2\kappa+1\big)\vD\times B=0.
$
In conclusion, the field equations are satisfied by the
ASD Ansatz, provided
\begin{equation}
     \kappa=-\half.
\end{equation}

Note that this solution, obtained again by using the
results in \cite{BakNO, LMSNO},
has nonvanishing electric field and also
a nonvanishing lower component,
namely $\chi=(i/2)D_{+}\Phi=iD_{1}\Phi=-D_{2}\Phi$.

\section{Conclusion}

In our previous paper \cite{HMS}, we found that a scalar field theory
augmented with a ``pure'' potential $V(\rho_{a})$, $a=l,r$ admitted
both the conventional and implementing the boosts from the left, also
the ``exotic'' Galilean symmetry. When a $U(1)_{\star}$ gauge field
with Chern-Simons is added, both these implementations are broken,
but Galilean symmetry
can be restored by having the boost act from the right.
Then we recover the exotic symmetry up to changing the sign ot the
noncommutative parameter $\theta$.
This is, furthermore, the unique implementation consistent with the
Seiberg-Witten map. It is remarkable that the interactions
determine the way Galilean symmetry should act.

Interestingly, the theory can be modified so that the fundamental
representation of the boosts, $\delta^l_{}$, acts as a symmetry. Switching
form the covariant derivative $D^l_{\mu}\equiv D_{\mu}$ in (\ref{lcovder}) to
\begin{equation}
     D^r_{\mu}\psi=\p_{\mu}\psi+\psi\star(ieA_{\mu})
     \label{rcovder}
\end{equation}
merely results in replacing the left-quantities
${\vj}_{}^{l}$ and $\rho^l$ in (\ref{ldenscur})
by {\it  minus} the corresponding right-quantities (\ref{rdenscur}).
As this latter transforms
homogeneously under the left-boost $\delta^l$ (and inhomogeneously under
the right-boost $\delta^r$),
\begin{equation}
     \delta^l\rho^r=-t\vb\cdot\vnabla\rho^r
     \qquad\hbox{and}\qquad
     \delta^r\rho^r=
     \theta\vb\times\vnabla\rho^r-t\vb\cdot\vnabla\rho^r,
     \label{lchangedens}
\end{equation}
cf. (\ref{fundchangedens}-\ref{afundchangedens}),
the new Gauss law with $\rho^r$
is form-invariant w. r. t. the fundamental representation
$\delta^l$. The invariance of the remaining equations can be shown readily.
Then the (right)boost (\ref{rncboost}) becomes left-boost,
\begin{equation}
     \vG^l=t\vP-\int\! d^2\vx\,\vx\rho^l=t\vP-
     \int\!x_{i}\vert\psi\vert^2\, d^2\vx
     +\frac{\theta}{2}\epsilon_{ij}\,\int\!\frac{1}{2i}
     \left(\bar{\psi}\p_{j}\psi-(\overline{\p_{j}\psi})\psi\right) d^2\vx.
     \label{lncboost}
\end{equation}
Switching form the right-handed expressions to the left-handed one
  amounts hence to changing the sign of $\theta$.

The second part of this paper is devoted to a discussion of various
vortex solutions.
First we studied spin $\half$ particles, described by the
L\'evy-Leblond equation. After demonstrating the (exotic) Galilean
symmetry, we have shown how spinning nontopological
vortices can be constructed using previous results.
Due to the breaking of the scale invariance cf. Section
\ref{scale},
we can not guarantee, however, that all solutions would be
self-dual, even for the critical value of the coupling.
Finally, we presented
  topological scalar vortices in a constant
(electro)magnetic background.
\goodbreak

\kikezd{Acknowledgement}.
Partial financial support is acknowledged to the PRIN SINTESI project of
the Italian Ministry of Education and Scientific Research.
One of us (P. H.) would like to thank
the University of Lecce for hospitality extended to him,
and to M. Hassa\"\i ne for discussions.
We are indebted to Fidel Schaposnik for enlightening
correspondence.

\end{document}